\documentclass{sig-alternate}
\usepackage{graphicx}
\usepackage{hyperref}
\usepackage{amsmath,amssymb}
\usepackage{subfig}
\usepackage{algorithm}
\usepackage{algorithmic}
\usepackage{multirow}
\usepackage{array}
\usepackage{balance}
\newtheorem{property}{Property}
\DeclareCaptionType{copyrightbox}

\newcommand{\DEMON}{\mbox{\it DEMON}}

\begin{document}
\conferenceinfo{KDD'12,} {August 12--16, 2012, Beijing, China.} 
\CopyrightYear{2012} 
\crdata{978-1-4503-1462-6 /12/08} 
\clubpenalty=10000 
\widowpenalty = 10000

\title{DEMON: a Local-First Discovery Method\\for Overlapping Communities}
\numberofauthors{2}

\author{
% 1st. author
\alignauthor
Michele Coscia\\
       \affaddr{CID - Harvard Kennedy School}\\
       \affaddr{79 JFK Street, Cambridge, MA, US}\\
       \email{michele\_coscia@hks.harvard.edu}
\alignauthor
Giulio Rossetti\\
       \affaddr{KDDLab University of Pisa}\\
       \affaddr{Largo B. Pontecorvo 3, Pisa, Italy}\\
       \email{giulio.rossetti@isti.cnr.it}
\and
% 2nd. author
\alignauthor
Fosca Giannotti\\
       \affaddr{KDDLab ISTI-CNR}\\
       \affaddr{Via G. Moruzzi 1, Pisa, Italy}\\
       \email{fosca.giannotti@isti.cnr.it}
% 3rd. author
\alignauthor
Dino Pedreschi\\
       \affaddr{KDDLab University of Pisa}\\
       \affaddr{Largo B. Pontecorvo 3, Pisa, Italy}\\
       \email{pedre@di.unipi.it}
}

\maketitle

\begin{abstract}
Community discovery in complex networks is an interesting problem with a number of applications, especially in the knowledge extraction task in social and information networks. However, many large networks often lack a particular community organization at a global level. In these cases, traditional graph partitioning algorithms fail to let the latent knowledge embedded in modular structure emerge, because they impose a top-down global view of a network. We propose here a simple local-first approach to community discovery, able to unveil the modular organization of real complex networks. This is achieved by democratically letting each node vote for the communities it sees surrounding it in its limited view of the global system, i.e. its ego neighborhood, using a label propagation algorithm; finally, the local communities are merged into a global collection. We tested this intuition against the state-of-the-art overlapping and non-overlapping community discovery methods, and found that our new method clearly outperforms the others in the quality of the obtained communities, evaluated by using the extracted communities to predict the metadata about the nodes of several real world networks. We also show how our method is deterministic, fully incremental, and has a limited time complexity, so that it can be used on web-scale real networks.
\end{abstract}
\vspace{-0.3cm}
\vspace{-0.3cm}

\section{Introduction}
Complex network analysis has emerged as one of the most exciting domains of data analysis and mining over the last decade. One of the most prolific sub field is community discovery in complex network, or {\em CD} in short. The concept of a ``community'' in a (web, social, or informational) network is intuitively understood as a set of individuals that are very similar, or close, to each other, more than to anybody else outside the community \cite{myreview}. This has often been translated in network terms into finding sets of nodes densely connected to each other and sparsely connected with the rest of the network. Community discovery can be seen as a network variant of traditional data clustering. To efficiently detect these structures is very useful for a number of applications, ranging from targeted vaccinations and outbreak prevention \cite{bionet}, to viral marketing \cite{viralm} and to many web data analysis tasks such as finding tribes in online information exchanges \cite{gurumine, dashun}, data compressing, clustering \cite{boldi} and sampling \cite{sampling}.

\begin{figure*}
\centering
\subfloat[A global view of the Facebook graph from 15k users.]{\includegraphics[scale=0.12]{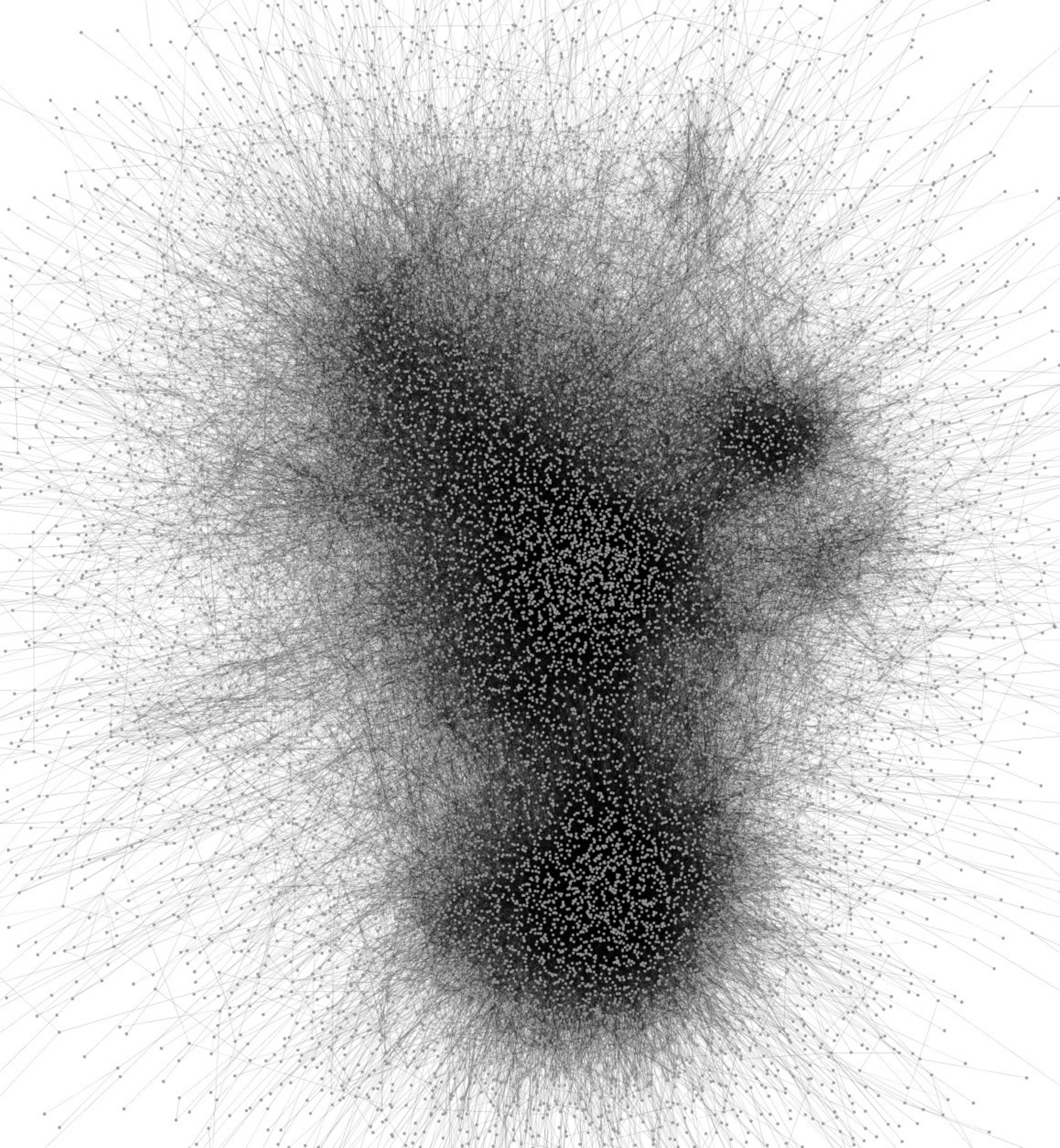}}\qquad\qquad\qquad\qquad
\subfloat[The ``ego minus ego'' network of one Facebook user among the 15k.]{\includegraphics[scale=0.24]{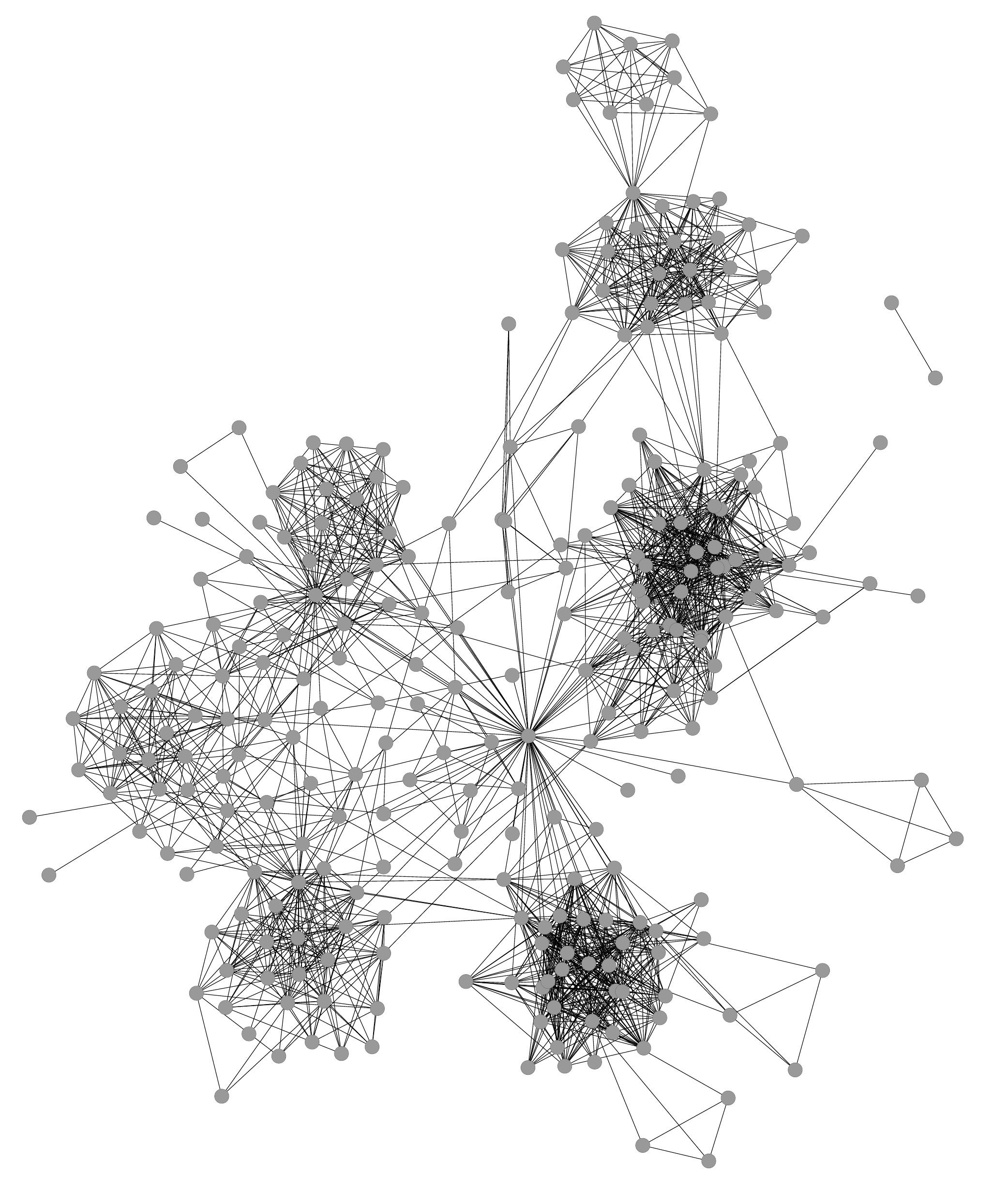}}
\caption{The real world example of the ``local vs global'' structure intuition.}
\label{fig:hairball-vs-ego}
\end{figure*} 

The classical problem definition of community discovery finds a very intuitive counterpart for small networks, where the denser areas are easily identifiable by visual inspection, while the problem becomes much harder for medium and large scale networks. At the global level, very little can be said about the modular structure of the network, because on larger scales the organization of the system becomes simply too complex. The friendship graph of Facebook includes more than 845 millions nodes as of February 2012\footnote{http://newsroom.fb.com/content/default.aspx?NewsAreaId=22}, but the difficulty of the CD task can be appreciated even considering a tiny fragment of the Facebook friendship graph, illustrated in Figure \ref{fig:hairball-vs-ego}(a). We depicted the connections among 15,000 nodes, i.e., less than 0.002\% of the total network. Even in this small subset of the network, no evident organization can be identified easily. Big networks are not analyzable with the naked eye. Very often, a visualization of ten thousands nodes results in a \textit{structureless} hairball. In cases like this, also generic community discovery algorithms tend to return not meaningful communities, as they typically try to cluster the whole structure and return some huge communities and a long list of small branches (see \cite{myreview}). Often, superimposing an order with a top-down approach leads to failure.

On the contrary, human eyes are good in finding denser areas in simple networks, i.e., the structure of cohesive groups of nodes that emerge considering a {\em local} fragment of an otherwise big network. But what does {\em local} mean? Commonsense goes that people are good at identifying the reasons why they know the people they know; therefore, each node has presumably an incomplete, yet clear, vision of the social communities it is part of, and that surrounds it. The consequences of exploiting this idea for the CD problem is effectively illustrated by Figure \ref{fig:hairball-vs-ego}(b). Here, we chose one of the 15k nodes from the previous example and extracted what we call its ``ego minus ego'' network, i.e. its ego network in which the ego node has been removed, together with all its attached edges. Suddenly, everything around the ego makes sense and some groups can be easily spotted. These groups correspond to the high school and university friends, mates from different workplaces and the members of an online community (we know all these details because the chosen ego is one of the authors of this paper). The ego is part of all these communities and knows that particular subsets of its neighborhood are part of these communities too. Probably, different egos have different perspectives over the same neighbors and it is the union of all these perspectives that creates an optimal partition of the network. In other words: if node $A$ and node $B$ are considered in the same communities by all the nodes connected to both $A$ and $B$, then they should be grouped in the same community. This is achieved by a {\em democratic} bottom-up mining approach: in turn, each node gives the perspective of the communities surrounding it and then all the different perspectives are merged together in an overlapping structure. 

In the vast CD literature, the general approach for the detection of the modular structure of a network is usually to develop a particular (greedy) algorithm, testing a general quality function with a particular heuristic and then return a set of communities extracted from the global structure (we discuss some of these methods in Section \ref{sec:related}). This approach generally fails for large networks due to the difference in structural organization at global and local scale. To cope with this difficulty, we propose a change of mentality. Since our community definition works perfectly in the small scale, then it should be applied only at this small scale. We propose a simple local-first approach to community discovery in complex networks by letting the hidden modular organization of a network emerge from local patterns.

Essentially, we adopt a {\em democratic} approach to the discovery of communities in complex networks. We ask each node to vote for the communities present in its local view of the network. For this reason, we chose to name our algorithm \textbf{D}emocratic \textbf{E}stimate of the \textbf{M}odular \textbf{O}rganization of a \textbf{N}etwork, or \DEMON\  in short. In practice, we extract the ego network of each node and apply a Label Propagation CD algorithm \cite{lp} on this structure, ignoring the presence of the ego itself, that will be judged by its peers neighbors. We then combine, with equity, the vote of everyone in the network. The result of this combination is a set of (overlapping) modules, the guess of the real communities in the global system, made not by an external observer, but by the actors of the network itself. Our democratic algorithm is {\em incremental}, allowing to recompute the communities only for the newly incoming nodes and edges in an evolving network. Nevertheless, \DEMON\  has also a low theoretical linear time complexity. The main core of our method has also the interesting property of being easily {\em parallelizable}, since only the ego network information is needed to perform independent computations, and it can be easily combined in a MapReduce framework \cite{mapreduce}; although the post-process Merge procedure is not trivially solvable in a MapReduce framework (and for this reason we leave a discussion about the parallel implementation as future work). The properties of \DEMON\  support its use in massive real world scenarios.

We provide an extensive empirical validation of \DEMON. In our experimental setting, we are particularly interested in investigating what useful knowledge we can discover. We test the results obtained with our method against selected state-of-the-art algorithms, both overlapping and not overlapping, since we believe that the possibility to cluster the same nodes in different communities is one of the crucial properties that a community discoverer should allow: online social networks have proved that individuals are part of many different communities and groups of interest. To evaluate this knowledge, we make use of a multilabel predictor fed with the extracted communities as input, with the aim of correctly classifying the metadata attached to the nodes in real life. Our datasets include the international store Amazon, the database of collaborations in movie industry IMDb, and the register of the activities of the US Congress GovTrack.us.

The rest of the paper is organized as follows: in Section \ref{sec:related} we present related works in community discovery literature. Section \ref{sec:problem} is dedicated to the problem representation and definition. Section \ref{sec:algo} describe the \DEMON\  algorithm structure, with algorithmic details and an account of the formal properties of the method. Our experiments are presented in Section \ref{sec:experiments}, and finally Section \ref{sec:conclusion} concludes the paper.

\section{Related Work}\label{sec:related}
The problem of finding communities in complex networks is very popular among network scientists, as witnessed by an impressive number of valid works in this field. A huge survey by Fortunato \cite{fortunato} explores all the most popular techniques to find communities in complex networks. Traditionally, a community is defined as a dense subgraph, in which the number of edges among the members of the community is significantly higher than the outgoing edges. However, this definition does not cover many real world scenarios, and in the years many different solutions started to explore alternative definitions of communities in complex networks \cite{myreview}.

A variety of CD methods are based on the {\em modularity} concept, a quality function of a partition proposed by Newman \cite{clauset-modularity, modularity}. Modularity scores high values for partitions in which the internal cluster density is higher than the external density. Hundreds of papers have been written about modularity, either using it as a quality function to be optimized, or studying its properties and deficiencies. One of the most advanced examples of modularity maximization CD is \cite{onnela}, where the authors use an extension of the modularity formula to cluster multiplex (evolving and/or multirelational) networks. A fast and efficient greedy algorithm, Modularity Unfolding, has been successfully applied to the analysis of huge web graphs of millions of nodes and billions of edges, representing the structure in a subset of the WWW \cite{mod-unfolding}.

Many algorithms have been proposed that are unrelated to modularity. Among them, a particular important field is the application of information theory techniques, as for example in Infomap \cite{infomap} or Cross Associations \cite{cct}. In particular, Infomap has been proven to be one among the best performing non overlapping algorithms \cite{lanci-comparative}. For this reason we chose Infomap as alternative to modularity approaches as a baseline method. Further, modularity approaches are affected by known issues, namely the resolution problem and the degeneracy of good solutions \cite{resolution-limit}. Similarly to Infomap, Walktrap \cite{walktrap} is based on flow methods and random walks.

A very important property for community discovery is the ability to return overlapping partitions, i.e., the possibility of a node to be part of more than one community. This property reflects the common sense intuition that each of us is part of many different communities, including family, work, and probably many hobby-related communities. Specific algorithms developed over this property are Hierarchical Link Clustering \cite{hlc}, HCDF \cite{hybrid} and k-clique percolation \cite{palla}.

Finally, an important approach is known as Label Propagation \cite{lp}: in this work authors detect communities by spreading labels through the edges of the graph and then labeling nodes according to the majority of the labels attached to their neighbors, iterating until a general consensus is reached. With a reasonable good quality on the partition, this algorithm is extremely fast and known to be one of the very few quasi-linear solutions to the community discovery problem, even if its plain application leads to worse results than Infomap and it does not return an overlapping partition. A related work is also \cite{bagrow}, whose aim is also to discover local communities. However, authors are only interested in those local communities and they do not return any global structure modular organization.

To to extract useful knowledge from the modular structure of networked data is also a prolific track of research. We recall the GuruMine framework, whose aim is to identify leaders in information spread and to detect groups of users that are usually influenced by the same leaders \cite{gurumine}. Many other works investigate the possibility of applying network analysis for studying, for instance, the dynamics of viral marketing \cite{viralm}.

\section{Networks and Communities}\label{sec:problem}
We model networks and their properties in terms of simple graphs.
For the sake of simplicity, a network is represented as an undirected, unlabeled and unweighted simple graph, denoted by $\mathcal{G}=(V,E)$ where $V$ is a set of nodes and $E$ is a set of edges, i.e., pairs $(u,v)$ representing the fact that there is a link in the network connecting nodes $u$ and $v$. It should be noted, however, that our method can handle weighted, directed and labeled multi-graphs.

In general terms, our problem definition is to find communities in complex networks. However, this is an ambiguous goal, as the definition itself of ``community'' in a complex network, similarly to the notion of clustering in statistics and data mining, is not unique \cite{myreview}. Furthermore, in a complex and semantically rich setting as the modern Web, one may want to cluster many different kinds of objects for many different reasons. Therefore, we need to narrow down our problem definition as follows.

We define two basic graph operations. The first one is the Ego Network extraction $EN$. Given a graph $\mathcal{G}$ and a node $v \in V$, $EN(v, \mathcal{G})$ is the subgraph $\mathcal{G}'(V', E')$, where $V'$ is the set containing $v$ and all its neighbors in $E$, and $E'$ is the subset of $E$ containing all the edges $(u, v)$ where $u \in V' \wedge v \in V'$. The second operation is the Graph-Vertex Difference $-g$: $-g(v, \mathcal{G})$ will result in a copy of $\mathcal{G}$ without the vertex $v$ and all edges attached to $v$. The combination of these two functions yields the $EgoMinusEgo$ function: $EgoMinusEgo(v, \mathcal{G}) = -g(v, EN(v, \mathcal{G}))$. Given a graph $\mathcal{G}$ and a node $v \in V$, the set of {\em local communities} $\mathcal{C}(v)$ of node $v$ is a set of (possibly overlapping) sets of nodes in $EgoMinusEgo(v, \mathcal{G})$, where each set $C \in \mathcal{C}(v)$ is a community according to node similarity: each node in $C$ is more similar to any other node in $C$ than to any other node in $C' \in \mathcal{C}(v)$, with $C \neq C'$. Finally, we define the set of {\em global communities}, or simply communities, of a graph $\mathcal{G}$ as:

\begin{equation}\label{eq:1}
\mathcal{C} =  Max ( \bigcup_{v \in V}\mathcal{C}(v) )
\end{equation}

where, given a set of sets $\mathcal{S}$, $Max(\mathcal{S})$ denotes the subset of $\mathcal{S}$ formed by its maximal sets only; namely, every set $S \in \mathcal{S}$ such that there is no other set $S' \in \mathcal{S}$ with $S \subset S'$. In other words, by equation (\ref{eq:1}) we generalize from local to global communities by selecting the maximal local communities that cover the entire collection of local communities, each found in the $EgoMinusEgo$ network of each individual node.

\section{The Algorithm}\label{sec:algo}
In this section we present our solution to the community discovery problem. The pseudo code of \DEMON\ is specified in Algorithm \ref{alg:demon}.

\subsection{The Core of the Algorithm}\label{sec:algo-core}

\begin{algorithm}[t]
\small
\begin{algorithmic}[1]
\REQUIRE{$\mathcal{G}:(V,E); \ \mathcal{C} = \emptyset; \epsilon  \in \left[0..1\right]$}
\ENSURE{set of overlapping communities $\mathcal{C}$}
\smallskip
\FORALL{$v\ \in\ V$}
\STATE $e\leftarrow EgoMinusEgo(v, \mathcal{G})$
\STATE $\mathcal{C}(v) \leftarrow LabelPropagation(e)$
\FORALL{$C\ \in\ \mathcal{C}(v)$}
\STATE $C  \leftarrow C \cup v$
\STATE $\mathcal{C} \leftarrow Merge(\mathcal{C},C,\epsilon)$
\ENDFOR
\ENDFOR 

\STATE {\bf return} $\mathcal{C}$
\end{algorithmic}
\caption{The pseudo-code of \DEMON\  algorithm.}
\label{alg:demon}
\end{algorithm}

The set of discovered communities $\mathcal{C}$ is initially empty. The external (explicit) loop of \DEMON\  cycles over each individual node, and it is necessary to generate all the possible points of view of the structure and get a complete coverage of the network itself. For each node $v$, we apply the $EgoMinusEgo(v, \mathcal{G})$ operation defined in Section \ref{sec:problem}, obtaining a graph $e$. We cannot apply simply the ego network extraction $EN(v, \mathcal{G})$ because the ego node $v$ is directly linked to all nodes $\in EN(v, \mathcal{G})$. This would lead to noise in the subsequent steps of \DEMON , since by our definition of local community the nodes would be put in the same community if they are close to each other. Obviously a single node connecting the entire sub-graph will make all nodes very close, even if they are not in the same community. For this reason, we remove the ego from its own ego network.

Once we have the $e$ graph, the next step is to compute the communities contained in $e$. We chose to perform this step by using a community discovery algorithm borrowed from the literature. Our choice fell on the Label Propagation (LP) algorithm \cite{lp}. This choice has been made for the following reasons:

\begin{enumerate}
\item LP  shares with this work the definition of what is a community.
\item LP  is known as the least complex algorithm in the literature, reaching a quasi-linear time complexity in terms of nodes. However,
\item LP  will return results of a quality comparable to more complex algorithms \cite{myreview}.
\end{enumerate}

Reason \#2 is particularly important, since Step \#3 of our pseudo code needs to be performed once for every node of the network. It is unacceptable to spend a superlinear time for each node at this stage, if we want to scale up to millions of nodes and hundreds of millions edges. Given the linear complexity of Step \#3, we refer to this as the internal (implicit) loop for finding the local communities.

We briefly describe in more detail the LP algorithm, given its importance in the \DEMON\  algorithm, following the original article \cite{lp}. Suppose that a node $v$ has neighbors $v_{1}, v_{2}, ... , v_{k}$ and that each neighbor carries a label denoting the community that it belongs to. Then $v$ determines its community based on the labels of its neighbors. A three-step example of this principle is shown in Figure \ref{fig:labelprop}. The authors assume that each node in the network chooses to join the community to which the maximum number of its neighbors belong. As the labels propagate, densely connected groups of nodes quickly reach a consensus on a unique label. At the end of the propagation process nodes with the same labels are grouped together as one community. Clearly, a node with an equal maximum number of neighbors in two or more communities can belong to both communities, thus identifying possible overlapping communities. The original algorithm does not handle this situation. For clarity, we report here the procedure of the LP algorithm, that is the expansion of Step \#3 of Algorithm \ref{alg:demon} and represents our inner loop:

\begin{figure}
\centering
\includegraphics[scale=0.185]{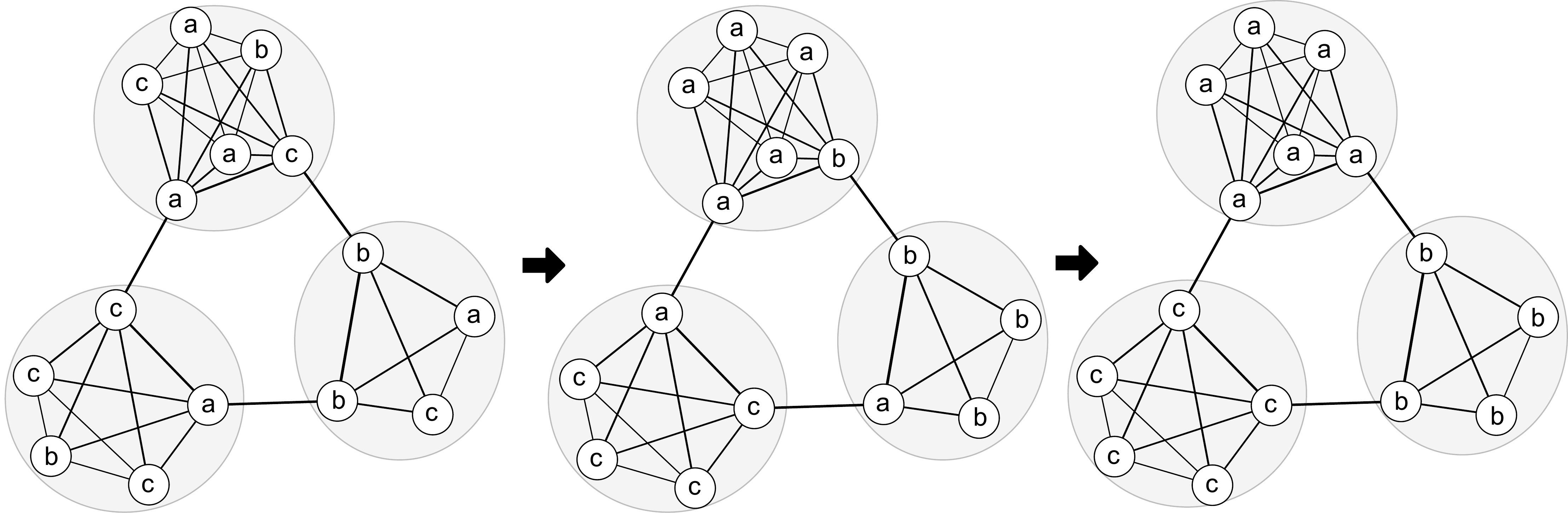}
\caption{A simple simulation of the Label Propagation process for community discovery.}\label{fig:labelprop}
\end{figure}

\begin{enumerate}
\item Initialize the labels at all nodes in the network. For any given node $v$, $C_v(0) = v$.
\item Set $t = 1$.
\item Arrange the nodes in the network in a random order and set it to $V$.
\item For each $v_i \in V$, in the specific order, let $C_{v_i}(t) = f(C_{v_{i1}}(t-1),  \ldots , C_{v_{ik}} (t-1))$. $f$ here returns the label occurring with the highest frequency among neighbors and ties are broken uniformly randomly.
\item If every node has a label that the maximum number of their neighbors have, or $t$ hits a maximum number of iterations $t_{max}$ then stop the algorithm. Else, set $t = t + 1$ and go to (3).
\end{enumerate}

\subsection{The Merge Function}\label{sec:algo-merge}
The result of Step \#3 of Algorithm \ref{alg:demon} is a set of local communities, according to the perspective of node $v$: at the end of the LP algorithm we reintroduce, in each local community, the node $v$. 
These communities are likely to be incomplete and should be used to enrich what \DEMON\  already discovered so far. Thus, the next step is to merge each local community of $\mathcal{C}$ in order to obtain the result set. The $Merge$ operation is defined as follows.

%$$Merge(c, \mathcal{C},\epsilon) = \left\{ 
%\begin{array}{lcl}
%\mathcal{C} & \quad & \exists c' \in \mathcal{C}: c \subseteq_{\epsilon} c' \\
% \{c\} \cup \{c' \in \mathcal{C} \mid c' \not\subseteq c \} & \quad & \mbox{otherwise}
%\end{array}
%\right.
%$$

\begin{algorithm}[t]
\small
\begin{algorithmic}[1]
\REQUIRE{$\mathcal{C} = Community\ set; C = Community; \epsilon \in \left[0..1\right]$}
\ENSURE{set of overlapping communities $\mathcal{C}$}
\smallskip
\FORALL{$ I\ \in\ \mathcal{C}$}
\IF{$C.size \le I.size and C\subseteq_{\epsilon}I$}
\STATE $u = C\cup I; $
\STATE $\mathcal{C}-C; \mathcal{C}-I;$
\STATE $\mathcal{C} = \mathcal{C} \cup u;$
\ENDIF

\ENDFOR 
\STATE {\bf return} $\mathcal{C}$
\end{algorithmic}
\caption{The pseudo-code of Merge  function.}
\label{alg:merge}
\end{algorithm}

Two communities $C$ and $I$ are merged if and only if  at most the $\epsilon\%$ of the smaller one is not included in the bigger one;
in this case, $C$ and $I$ are removed from $\mathcal{C}$  and their union is added to the result set.
The $\epsilon$ factor is introduced to vary the percentage of  common elements provided from each couple of communities: $\epsilon=0$ ensure that two communities
are merged only if one of them is a proper subset of the other, on the other hand with a value of  $\epsilon=1$ even communities that do not share a single node
are merged together.
  
% it is not covered by any community already in $\mathcal{C}$; in this case, all communities in $\mathcal{C}$ covered by $c$, if any, are removed. 
\subsection{DEMON Properties}\label{sec:algo-properties}
To prove the correctness of the \DEMON\  algorithm w.r.t.\ the problem definition in Section \ref{sec:problem}, we prove by induction that the following holds:

\begin{property}\label{eq:2} At the $k$-th iteration of the outer loop of \DEMON, for all $k \geq 0$:
\begin{equation}
\mathcal{C} = Max ( \bigcup_{v = v_1, \ldots, v_k}\mathcal{C}(v) )
\end{equation}
where $v_1, \ldots, v_k$ are the nodes visited after $k$ iterations. 
\end{property}
Property (\ref{eq:2}) trivially holds for $k=0$, i.e., at initialization stage. For $k > 0$, assume that the property holds up to $k-1$. Then $\mathcal{C}$ contains the maximal local communities of the subgraph with nodes ${v_1, \ldots, v_{k-1}}$. By merging every local community $C$ of node $v_k$ into $\mathcal{C}$, we guarantee that $C$ is added to the result only if it is not covered by any preexisting community, and, if added, any preexisting community covered by $C$ is removed from $\mathcal{C}$. As a result, after merging all communities in $\mathcal{C}(v_k)$ into $\mathcal{C}$ in Steps \#4-6, the latter is the set of maximal communities covering all local communities discovered in ${v_1, \ldots, v_{k}}$. Therefore, we can conclude that \DEMON\  is a correct and complete implementation of the CD problem stated by equation (\ref{eq:1}). More generally, denoting by $\DEMON(\mathcal{G}, \mathcal{C})$ the set of communities $\mathcal{C}'$ obtained by running the \DEMON\ algorithm on graph $\mathcal{G}$ starting with the (possibly non-empty) set of communities $\mathcal{C}$, the following properties hold.

\begin{property}\label{eq:correcteness} {\bf Correctness} and {\bf Completeness}.\\ 
If $\DEMON(\mathcal{G}, \mathcal{C}) = \mathcal{C}'$, where $\mathcal{G} = (V, E)$, then 
\begin{equation}
\mathcal{C}' = Max ( \mathcal{C} \cup \bigcup_{v \in V}\mathcal{C}(v) )
\end{equation}
\end{property}
In other words, given a preexisting set of communities $\mathcal{C}$ and a graph $\mathcal{G}$, \DEMON\ returns all and only the communities obtained extending $\mathcal{C}$ with the communities found in $\mathcal{G}$, coherently with the definition of communities given in equation (\ref{eq:1}).

\begin{property}
{\bf Determinacy} and {\bf Order insensitivity}.\\ 
There exists a {\em unique}\ $\mathcal{C}' = \DEMON(\mathcal{G}, \mathcal{C})$ for any given $\mathcal{G}$ and $\mathcal{C}$, disregarding the order of visit of the nodes in $\mathcal{G}$. 
\end{property}
This is a direct corollary of property (\ref{eq:correcteness}) and of the uniqueness of the set $Max(\mathcal{S})$ for any set of sets $\mathcal{S}$, under the assumption that the set of local communities $\mathcal{C}(v)$ is also uniquely assigned, for any node $v$. Therefore, the order in which the nodes in $\mathcal{G}$ are visited by \DEMON\ is irrelevant.

\begin{property}
{\bf Compositionality}. Consider any partition of a graph $\mathcal{G}$ into two subgraphs $\mathcal{G}_1$, $\mathcal{G}_2$ such that, for any node $v$ of $\mathcal{G}$, the entire ego network of $v$ in $\mathcal{G}$ is fully contained either in $\mathcal{G}_1$ or $\mathcal{G}_2$. Then, given an initial set of communities $\mathcal{C}$:
\begin{equation}\label{eq:compositionality}
\DEMON(\mathcal{G}_1 \cup \mathcal{G}_2, \mathcal{C}) = Max (\DEMON(\mathcal{G}_1, \mathcal{C}) \cup \DEMON(\mathcal{G}_2, \mathcal{C}))
\end{equation}  
\end{property}
This is a consequence of two facts: $i)$  each local community $\mathcal{C}(v)$ is correctly computed under the assumption that the subgraphs do not split any ego network, and $ii)$ for any two sets of sets $\mathcal{S}_1, \mathcal{S}_2$, $Max(\mathcal{S}_1 \cup \mathcal{S}_2) = Max(Max(\mathcal{S}_1) \cup Max(\mathcal{S}_2))$.

\begin{property}
{\bf Incrementality}. Given a graph $\mathcal{G}$, an initial set of communities $\mathcal{C}$ and an incremental update $\Delta \mathcal{G}$ consisting of new nodes and new edges added to $\mathcal{G}$, where $\Delta \mathcal{G}$ contains the entire ego networks of all new nodes and of all the preexisting nodes reached by new links, then
\begin{equation}\label{eq:incrementality}
\DEMON(\mathcal{G} \cup \Delta \mathcal{G}, \mathcal{C}) = \DEMON(\Delta \mathcal{G}, \DEMON(\mathcal{G}, \mathcal{C}))
\end{equation}
\end{property}
This is a consequence of the fact that only the local communities of nodes in $\mathcal{G}$ affected by new links need to be reexamined, so we can run \DEMON\ on $\Delta \mathcal{G}$ only, avoiding to run it from scratch on $\mathcal{G} \cup \Delta \mathcal{G}$. 

Properties (\ref{eq:compositionality}) and (\ref{eq:incrementality}) have important computational repercussions. The compositionality property entails that the core of \DEMON\ algorithm as described in subsection \ref{sec:algo-core} is highly parallelizable, because it can run independently on different fragments of the overall network with a relatively small combination work. Each node of the computer cluster needs to obtain a small fragment of the network, as small as the ego network of one or a few nodes. The Map function is simply the LP algorithm. The incrementality property entails that \DEMON\ can efficiently run in a streamed fashion, considering incremental updates of the graph as they arrive in subsequent batches; essentially, incrementality means that it is not necessary to run \DEMON\ from scratch as batches of new nodes and new links arrive: the new communities can be found by considering only the ego networks of the nodes affected by the updates (both new nodes and old nodes reached by new links). This does not trivially hold for the Merge function presented in subsection \ref{sec:algo-merge}, therefore the actual parallel implementation of \DEMON\ is left as future work. However, different and simpler Merge functions can be define to combine the results provided by the core of the algorithm, thus preserving its possibility to scale up in a parallel framework.

\subsection{Complexity}
We now evaluate the time complexity of our approach. \DEMON\ core (Section \ref{sec:algo-core}) is based on the Label Propagation algorithm, whose complexity is $\mathcal{O}(n + m)$ \cite{lp}, where $n$ is the number of nodes and $m$ is the number of edges. LP is performed once for each node. Let us assume that we are working with a scale free network, whose degree distribution is $p_k = k^{-\alpha}$. This means that there are $\frac{n}{k^\alpha}$ nodes with degree $k$. If $K$ is the maximum degree, the complexity would be $\sum_{k = 1}^{K}(\frac{n}{k^\alpha} \times (k + \frac{k(k -1)}{2}))$ because for each node of degree $k$ we have an ego network of $k$ nodes and at worst $\frac{k(k -1)}{2}$ edges. This number is very small for the vast majority of nodes, being the degree distribution right skewed, thus many nodes have degree 1 or 2. We omit the solution of the sum with the integral and we report that the complexity is then dominated by a single term, ending up to be $\mathcal{O}(nK^{3-\alpha})$. This means that the stronger is the $\alpha$ exponent, the faster is \DEMON : with $\alpha = 3$ we have few super-hubs for which we basically check the entire network few times and the rest of nodes add nothing to the complexity; with $\alpha = 2$ we have many high degree nodes and we end up with higher complexity, but still subquadratic in term of nodes (as, with $\alpha = 2$, $K << n$).

\section{Experiments}\label{sec:experiments}
We now present our experimental findings. We make use of three networked datasets, representing very different phenomena. We first concentrate on evaluating the quality of a set of communities discovered in these datasets, comparing the results with those of other competing methods in terms of the predictive power of the discovered communities. Since real world data are enriched with annotated information, we measure the ability of each community to predict the semantic information attached with the metadata of the nodes within the community itself.

Next, we assess the community quality using a global measure of community cohesion, based on the intuition that nodes into the same community should possess similar semantic properties in terms of attached metadata. 

The selected competitors for our assessment are: Hierarchical Link Clustering (HLC) \cite{hlc}, that has been proven able to outperform all the overlapping algorithms, including the $k$-clique Propagation algorithm by Palla et al \cite{palla}; two random walks based methods, one focusing on minimizing random walk entropy (Infomap \cite{infomap}) and the other relying on a general flow method (Walktrap \cite{walktrap}); a leading eigenvector-based community discovery, namely Modularity maximization in the fast greedy implementation introduced in \cite{clauset-modularity}. Finally, we present some examples of knowledge that we are able to extract from the communities found by the \DEMON\  algorithm.

Note that we are not able to provide the analytic evaluation for Amazon dataset: for that network HLC algorithm was not able to provide results due to memory consumption problems, while the other community discovery algorithms usually returned some huge communities that was not possible to analyze (see Section \ref{sec:prediction} and particularly Figure \ref{fig:sizedistr} for more information).

The experiments were performed on a Dual Core Intel i7 64 bits @ 2.8 GHz, equipped with 8 GB of RAM and with a kernel Linux 3.0.0-12-generic (Ubuntu 11.10). The code was developed in Java and it is available for download with the network datasets used\footnote{http://www.di.unipi.it/$\sim$coscia/demon/}. For performances purposes, we mainly refer to the biggest dataset, i.e. Amazon: the core of the algorithm (Section \ref{sec:algo-core}) took less than a minute, while the Merge function (Section \ref{sec:algo-merge}) with increasing thresholds can take from one minute to one hour.

\subsection{Networks}
We tested our algorithms on three real world complex networks extracted from available web services of different domains. A general overview about the statistics of these networks can be found in Table \ref{tab:networks}, where: $|V|$ is the number of nodes, $|E|$ is the number of edges and $\bar{k}$ is the average degree of the network. Congress and IMDb networks are similar to the ones used in \cite{hlc}, generally updating the source dataset with a more recent set of data, and we refer to that paper for a deeper description of them. The networks were generated as follows:

\textbf{Congress}. The network of legislative collaborations between US representatives of the House and the Senate during the 111st US congress (2009-2011). We downloaded the data about all the bills discussed during the last Congress from GovTrack\footnote{http://www.govtrack.us/developers/data.xpd}, a web-based service recording the activities of each member of the US Congress. The bills are usually co-sponsored by many politicians. We connect politicians if they have at least 75 co-sponsorships and delete all the connections that are created only by bills with more than 10 co-sponsors. Attached to each bills in the Govtrack data we have also a collection of subjects related to the bill. The set of subjects a politicians frequently worked on is the \textit{qualitative attribute} of this network.

\textbf{IMDb}. We downloaded the entire database of IMDb from their official APIs\footnote{http://www.imdb.com/interfaces} on August 25th 2011. We focus on actors who star in at least two movies during the years from 2001 to 2010, filtering out television shows, video games, and other performances. We connect actors with at least two movies in which they both appear. This network is weighted according to the number of co-appearances. Our \textit{qualitative attributes} are the user assigned keywords, summarizing the movies each actor has been part of.

\textbf{Amazon}. We downloaded Amazon data from the Stanford Large Network Dataset Collection\footnote{http://snap.stanford.edu/data/index.html}. In this dataset, frequent co-purchases of products are recorded for the day of May 5th 2003. We transformed the directed network in an undirected version. We also downloaded the metadata information about the products, available in the same repository. Using this metadata, we can define the \textit{qualitative attributes} for each product as its categories.

\begin{table}
\centering
\begin{tabular}{|l|rrr|}
\hline
Network & $|V|$ & $|E|$ & $\bar{k}$ \\
\hline
Congress & 526 & 14,198 & 53.98 \\
%Philosophers & 1,260 & 6,739 & 10.69 \\
IMDb & 56,542 & 185,347 & 6.55 \\
Amazon & 410,236 & 2,439,437 & 11.89 \\
\hline 
\end{tabular}
\caption{Basic statistics of the studied networks.}
\label{tab:networks}
\end{table}

\subsection{Quality Evaluation via Label Prediction}\label{sec:prediction}
We first assess \DEMON\ performances using a classical prediction task. We attach the community memberships of a node as known attributes, then its qualitative attributes (real world labels) as target to be predicted; we then feed these attributes to a state-of-the-art label predictor and record its performance. Of course, a node may have one or more known attributes, as both \DEMON\ and HLC are overlapping community discoverers; and it may have also one or more unknown attributes, as it can carry many different labels.

For this reason, we need a multilabel classificator, i.e. a learner able to predict multiple target attributes \cite{brl}. We chose to use the Binary Relevance Learner. The BRL learns $|L|$ binary classifiers $H_l:X \rightarrow \{l, \neg l\}$, one for each different label $l \in L$. It transforms the original data set into $|L|$ data sets $D_l$ that contain all examples of the original data set, labeled as $l$ if the labels of the original example contained $l$ and as $\neg l$ otherwise. It is the same solution used in order to deal with a single-label multi-class problem using a binary classifier. Note that this classifier does not penalize per se non-overlapping partitions, as each target label is classified independently, and this property is requested to fairly confront overlapping algorithms such as \DEMON\ and HLC, with the other non-overlapping algorithms. We used the Python implementation provided in the Orange software\footnote{http://orange.biolab.si/}. For time and memory constraints due to the BRL complexity, for IMDb we used as input only the biggest communities (with more than 15 nodes) and eliminating all nodes that are not part of any of the selected communities.

Multi-label classification requires different metrics than those used in traditional single-label classification. Among the measures that have been proposed in the literature, we use the multi-label version of the standard Precision and Recall measures. Let $D_l$ be our multi-label evaluation data set, consisting of $|D_l|$ multi-label examples $(x_i,Y_i), i=1..|D_l|, Y_i \subseteq L$. Let $H$ be our BRL multi-label classifier and $Z_i=H(x_i)$ be the set of labels predicted by $H$ for $x_i$. Then, we can evaluate Precision and Recall of $H$ as:

$$Precision(H,D_l)=\frac{1}{|D_l|} \sum_{i=1}^{|D_l|} \frac{|Y_i \cap Z_i|}{|Z_i|},$$

$$Recall(H,D_l)=\frac{1}{|D_l|} \sum_{i=1}^{|D_l|} \frac{|Y_i \cap Z_i|}{|Y_i|}.$$

\begin{table}
\centering
\tiny
\begin{tabular}{|l|rrrrr|}
\hline
Network & \DEMON & HLC & Infomap & Modularity & Walktrap \\
\hline
Congress & \textbf{0.21275} & 0.14740 & 0.00535 & 0.00099 & 0.00725 \\
IMDb & \textbf{0.44252} & 0.43078 & 0.38470 & 0.10692 & 0.17488 \\
\hline
\end{tabular}
\caption{The F-Measure scores for Congress and IMDb dataset and each community partition.} \label{tab:fmeasure}
\end{table}

\begin{table*}
\centering
\begin{tabular}{|l|rr|rr|rr|rr|rr|}
\hline
\multirow{2}{*}{Network} & \multicolumn{2}{c|}{Demon} & \multicolumn{2}{c|}{HLC} &  \multicolumn{2}{c|}{Infomap} &  \multicolumn{2}{c|}{Modularity} &  \multicolumn{2}{c|}{Walktrap}\\
& $|\mathcal{C}|$ & $\bar{|c|}$ &  $|\mathcal{C}|$ & $\bar{|c|}$ & $|\mathcal{C}|$ & $\bar{|c|}$ & $|\mathcal{C}|$ & $\bar{|c|}$ & $|\mathcal{C}|$ & $\bar{|c|}$\\
\hline
Congress & 425 & 63.3671 & 1,476 & 4.5867 & 6 & 87.6667 & 3 & 175.3333 & 7 & 71.8571 \\  
IMDb & 14,004 & 12.6824 & 88,119 & 8.3426 & 5,991 & 27.1574 & 4,746 & 11.9157 & 7,877 & 7.1781 \\  
\hline
\end{tabular}
\caption{Statistics of the community set returned by the different algorithms.}
\label{tab:commset-stats}
\end{table*}

We then derive the F-measure from Precision and Recall. For alternatives multi-label evaluations, we refer to \cite{mleval-pakdd}. The results are reported in Table \ref{tab:fmeasure} and show how \DEMON\   outperforms its competitors. We did not test Amazon network as HLC was not able to provide results due to its complexity and further the BRL classifier was not able to scale for the overall number of nodes and labels.

For IMDb dataset, HLC was able to score almost like \DEMON . However, there is an important distinction to be made about the quantity of the results: if the community discovery returns too many communities, then it is difficult to actually extract useful knowledge from them. We reported in Table \ref{tab:commset-stats} the basic statistics about the partitions returned by the algorithms: number of communities ($|\mathcal{C}|$) and average community size ($\bar{|c|}$). For \DEMON , we report the statistics of the communities extracted with $\epsilon = 0$. As we can see, not only  \DEMON\ scores better results, but it does with 70-80\% less communities than HLC and with an average community size more manageable than Infomap.

\begin{figure}
\centering
\subfloat[Congress]{\includegraphics[width=.5\columnwidth]{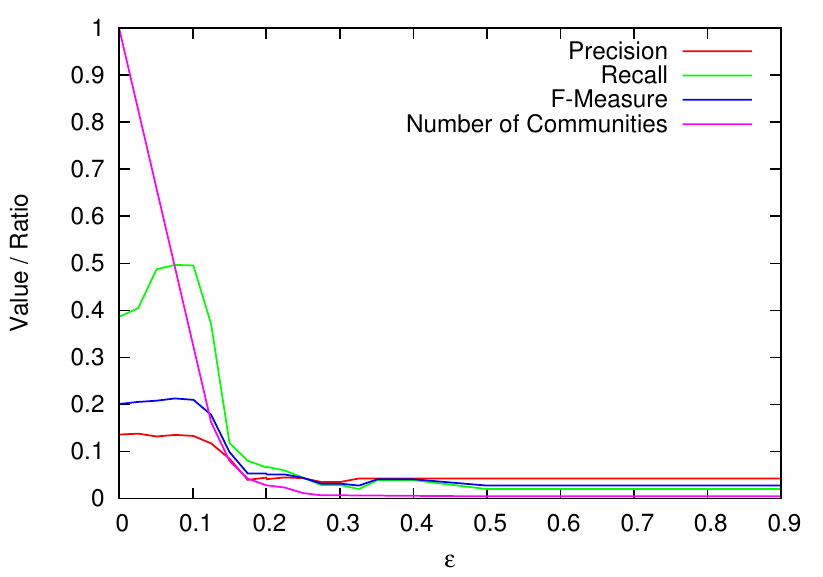}}
\subfloat[IMDb]{\includegraphics[width=.5\columnwidth]{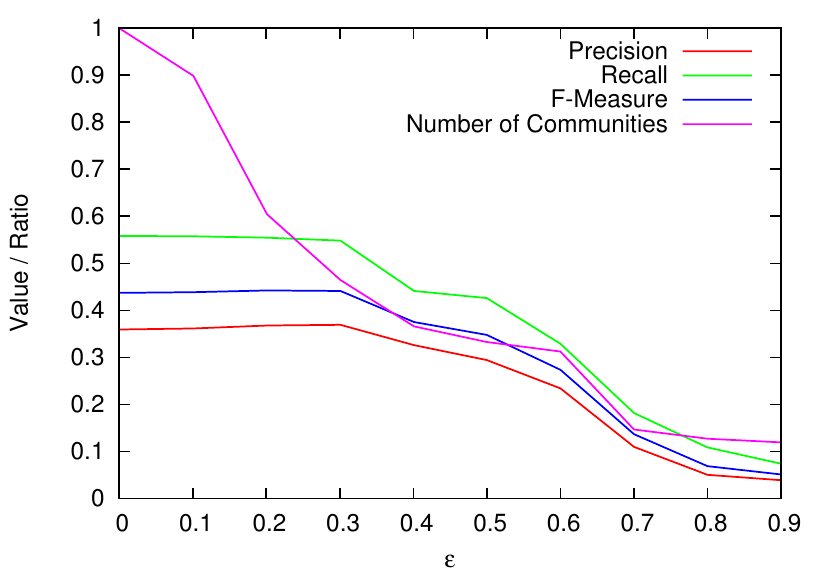}}
\caption{Precision, Recall, F-Measure and number of communities for different $\epsilon$ values.}\label{fig:commset-stats}
\end{figure}

We report in Table \ref{tab:commset-stats} the results for $\epsilon = 0$. However, we vary the $\epsilon$ threshold and see what happens to the number of communities and to the quality of the results. We can see that for both Congress and IMDb the Precision, Recall and F-Measure scores remains constant (and actually F-Measure peaks at $\epsilon = 0.076$ and $\epsilon = 0.301$ for Congress and IMDb respectively) before falling for increasing $\epsilon$ values, while the relative number of communities dramatically drops. For Congress, we have the maximum F-Measure with only 175 communities; while for IMDb F-Measures peaks with 6,508 communities (in both cases, less than 50\% of the communities at $\epsilon = 0$ and than an order of magnitude of HLC).

\begin{figure}
\centering
\includegraphics{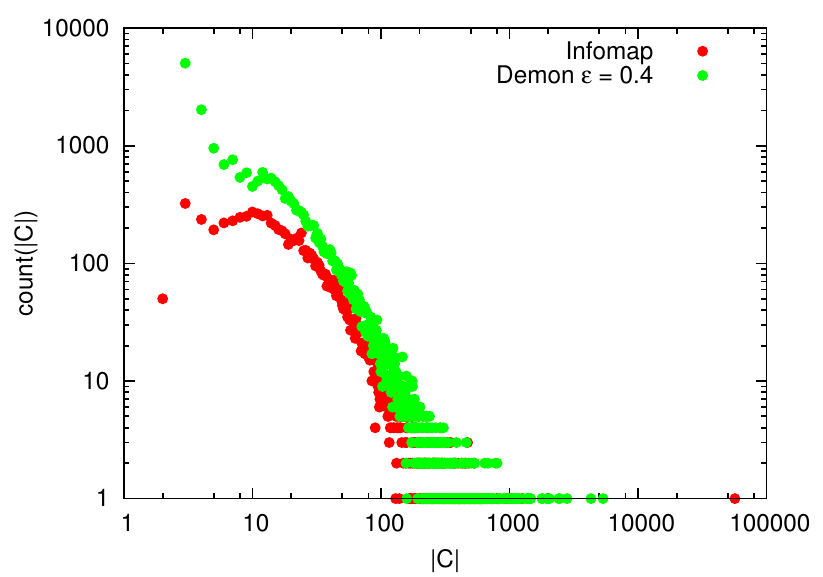}
\caption{The distribution of the community sizes for DEMON and Infomap in the Amazon network.}\label{fig:sizedistr}
\end{figure}

A final consideration is needed about the size distribution of the communities detected by \DEMON\ and the other community discovery algorithms. In Figure \ref{fig:sizedistr} we depicted the community size distribution for \DEMON\ and Infomap for the Amazon network. As we can see, Infomap returned, among the others, a giant community, one order of magnitude greater than the biggest one returned by \DEMON . To analyze such a community has been proved impossible, and this is another reason why we are not able to provide an analitycal evaluation of the results extracted from the Amazon network.

We can conclude that \DEMON\ with a manageable number of communities is able to outperform more complex methods and the choice of $\epsilon$ can make the difference in obtaining a narrower set of communities with the same (or greater) overall quality.

\subsection{Quality Evaluation via Community Cohesion}
As presented at the beginning of this section, the networks studied here possess \textit{qualitative attributes} that attaches a small set of annotations or tags to each node. Assuming that these qualitative attributes form a description of the node, beyond the network itself, we can reasonably state that ``similar'' nodes share more \textit{qualitative attributes} than dissimilar nodes. This procedure is not standard in community discovery results evaluation. Usually authors prefer to use the established measure of Modularity. However, Modularity is strictly (and exclusively) dependent on the graph structure. What we want evaluate is not how a graph measure is maximized, but how good is our community partition in describing real world knowledge about the clustered entities.

We quantify the matching between a community partition and the metadata by evaluating how much higher are on average the Jaccard coefficient of the set of \textit{qualitative attributes} for pair of nodes inside the communities over the average of the entire network, or:

$$
CQ(P) = \dfrac{\sum_{(n_1, n_2) \in P}\frac{|QA(n_1) \cap QA(n_2)|}{|QA(n_1) \cup QA(n_2)|}}{\sum_{(n_1, n_2) \in E}\frac{|QA(n_1) \cap QA(n_2)|}{|QA(n_1) \cup QA(n_2)|}},
$$

where $P$ is the set of node pairs that share at least one community, $QA(n)$ is the set of qualitative attributes of node $n$ and $E$ is the set of all edges. If $CQ(P) = 1$, then there is no difference between $P$ and the average similarity of nodes, i.e. $P$ is practically random. Lower values implies that we are grouping together dissimilar nodes, higher values are expected for an algorithm able to group together similar nodes.

To calculate the Jaccard coefficient for each pair of the network is computationally prohibitive. Therefore, for IMDb we chose a random set of 400k pairs. Moreover, CQ is biased towards algorithms returning more communities. For this reason, we just collected random communities from the community pool, trying to avoid too much overlap as we want also to maximize the number of nodes considered by CQ (i.e. we try not to consider more than one community per node). We apply this procedure for each algorithm and calculated the CQ value. We repeated this process for 100 iterations and we report in Table \ref{tab:cq} the average value of the CQ obtained. Also in this case, \DEMON\ was able to easily outperform all the other algorithms.

\begin{table}
\centering
\tiny
\begin{tabular}{|l|rrrrr|}
\hline
Network & \DEMON & HLC & Infomap & Modularity & Walktrap \\
\hline
Congress & \textbf{1.1792} & 1.1539 & 1.0229 & 1.0373 & 1.0532 \\
IMDb & \textbf{5.6158} & 5.1589 & 0.1400 & 1.4652 & 0.0211 \\
\hline
\end{tabular}
\caption{The Community Quality scores for Congress and IMDb dataset and each community partition.} \label{tab:cq}
\end{table}

\subsection{Interpretation of Discovered Communities}
In this Section we present a brief case study using the communities extracted for the previously exposed evaluation of \DEMON . Aim of the section is to demonstrate that the extracted communities have practical applications in the extraction of knowledge from real world scenarios. In the Amazon network to have different communities for each item is very useful. A recommendation system is able to better discern if a user may be interested in a product or not given that he bought something else; however being part of one community of products does not mean that that particular community describes all aspects of a particular product.

\begin{figure}
\centering
\includegraphics[scale=0.33]{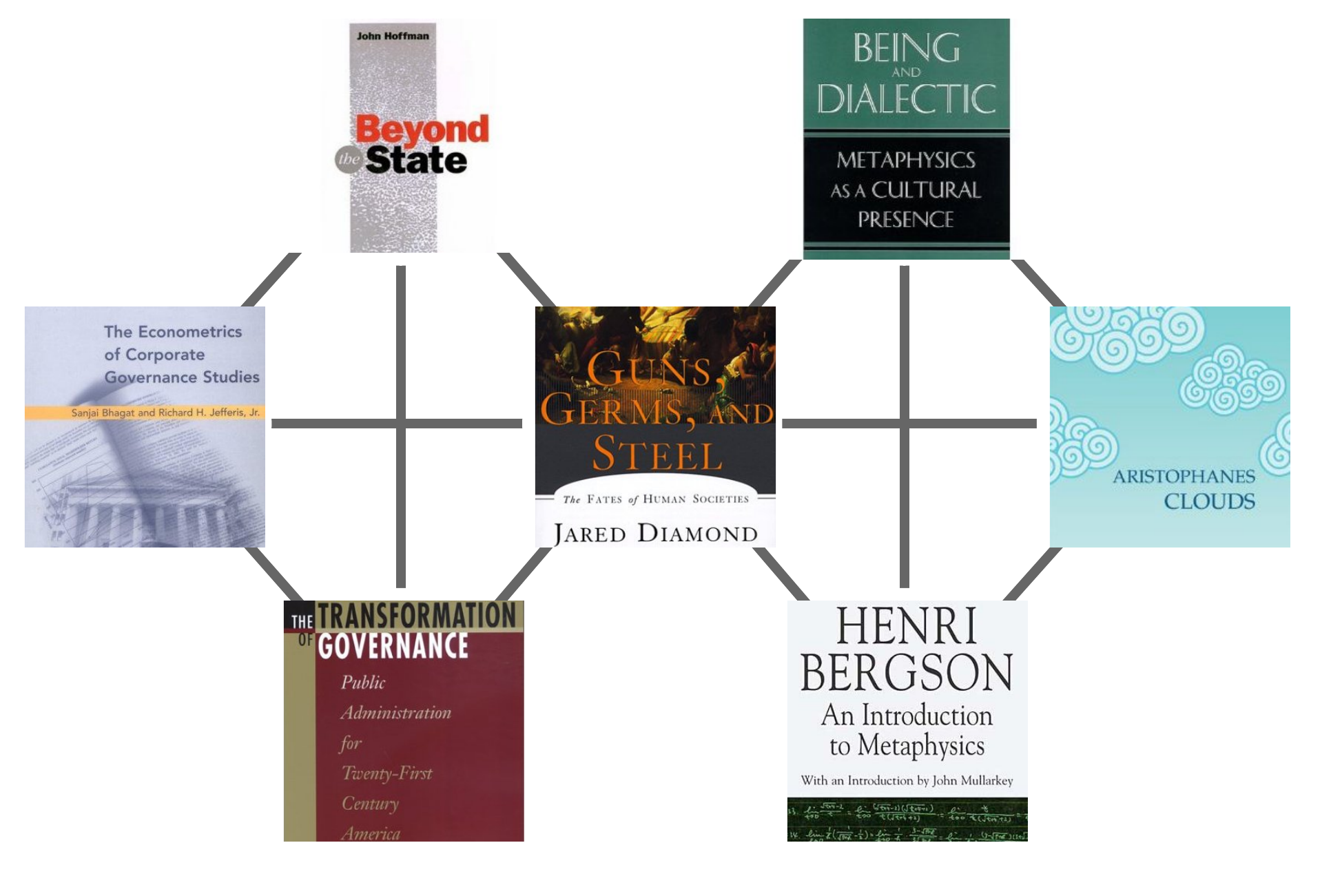}
\caption{A representation of parts of the two communities surrounding our case study in the amazon network.}\label{fig:amazon-examples}
\end{figure}

Let us consider, as an example, the case of Jared Diamond's best selling book ``Guns, Germs, and Steel: The Fates of Human Societies''. Clearly, it is difficult to say that the people interested in this book are always interested in the same things. Checking the communities to which it belongs, we find two very different big communities (a depiction of the two communities is provided by Figure \ref{fig:amazon-examples}). These communities have some sort of overlap, however they can be characterized by looking at the products that appear exclusively in one or in the other. In the first one we find books such as: ``Beyond the State: An Introductory Critique'', ``The Econometrics of Corporate Governance Studies'' and ``The Transformation of Governance: Public Administration for Twenty-First Century America''. This is clearly a community composed mainly by purchases made by the people more interested in the socioeconomic aspects of Diamond's book. The second community hosts products such as: ``An Introduction to Metaphysics'', ``Aristophanes' Clouds Translated With Notes and Introduction'' and ``Being and Dialectic: Metaphysics and Culture''. This second communities is apparently composed by the purchases of customers more attracted by the underlying philosophical implications of Diamond's study. Products in one communities may have something in common, but they are part of two distinct and very well characterized groups, and the one in one group are not expected to be found in the other.

This is of course one of the many cases. We report as an additional example the two communities around the historical novel ``The Name of the Rose'' by Umberto Eco: one community is characterized by history related products (such as ``Ancestral Passions : The Leakey Family and the Quest for Humankind's Beginnings''), the other by costume fiction (for example the 1932 Dreyer's movie ``Vampyr'').

\section{Conclusion and Future Works}\label{sec:conclusion}
In this paper we proposed a new method for solving the problem of detecting latent knowledge from significant communities in complex networks. We propose a democratic approach, where the peer nodes judge where their neighbors should be clustered together. This approach has robust theoretical properties, including correctness and completeness w.r.t.\ a precise community definition, determinacy, compositionality and incrementality, that make it amenable to challenge the conceptual and computational challenge to analyze networks with millions of nodes. We have shown in the experimental section that this method allows a discovery of communities in different real world networks collected from information rich datasets. The quality of the overlapping partition, a partition that allows nodes to be in different communities at the same time, is improved w.r.t state-of-the-art algorithms, evaluated using the communities to predict the metadata attached to the nodes, and according to a quantitative quality function, also metadata-based.

Many lines of research remain open for future work, such as an efficient parallel implementation that may make \DEMON\  the first community discovery algorithm able to scale to billions of nodes; different merging strategies that may further improve the quality of the results; different hosted algorithms can be used instead of the Label Propagation algorithm in the inner loop of \DEMON, to extract communities according to different definitions.\\

\textbf{Acknowledgments.} Michele Coscia is a recipient of the Google Europe Fellowship in Social Computing, and this research is supported in part by this Google Fellowship. This work has been partially supported by the European Commission under the FET-Open Project n. FP7-ICT-270833, DATA SIM -- DATa science for SIMulating the era of electric vehicles http://www.datasim-fp7.eu/.

\balance

\bibliographystyle{plain}

\bibliography{kddrt563-coscia-arxiv}

\begin{thebibliography}{10}

\bibitem{hlc}
Yong-Yeol Ahn, James~P. Bagrow, and Sune Lehmann.
\newblock {Link communities reveal multiscale complexity in networks}.
\newblock {\em Nature}, 466(7307):761--764, June 2010.

\bibitem{bagrow}
James~P. Bagrow and Erik~M. Bollt.
\newblock Local method for detecting communities.
\newblock {\em Physical Review E}, 72(4):046108+, October 2005.

\bibitem{mod-unfolding}
Vincent~D. Blondel, Jean-Loup Guillaume, Renaud Lambiotte, and Etienne
  Lefebvre.
\newblock Fast unfolding of communities in large networks.
\newblock {\em J.STAT.MECH.}, page P10008, 2008.

\bibitem{boldi}
Paolo Boldi, Marco Rosa, Massimo Santini, and Sebastiano Vigna.
\newblock Layered label propagation: a multiresolution coordinate-free ordering
  for compressing social networks.
\newblock In {\em WWW}, pages 587--596, 2011.

\bibitem{clauset-modularity}
Aaron Clauset, M.~E.~J. Newman, and Cristopher Moore.
\newblock Finding community structure in very large networks.
\newblock {\em Physical Review E}, 70:066111, 2004.

\bibitem{myreview}
Michele Coscia, Fosca Giannotti, and Dino Pedreschi.
\newblock A classification for community discovery methods in complex networks.
\newblock {\em Statistical Analysis and Data Mining}, 4(5):512--546, 2011.

\bibitem{mapreduce}
Jeffrey Dean and Sanjay Ghemawat.
\newblock {MapReduce}: Simplified data processing on large clusters.
\newblock {\em OSDI}, pages 137--150, 2004.

\bibitem{palla}
Imre Der\'{e}nyi, Gergely Palla, and Tam\'{a}s Vicsek.
\newblock {Clique Percolation in Random Networks}.
\newblock {\em Physical Review Letters}, 94(16):160202+, April 2005.

\bibitem{fortunato}
S.~{Fortunato}.
\newblock {Community detection in graphs}.
\newblock {\em Physics Reports}, 486:75--174, February 2010.

\bibitem{resolution-limit}
Santo Fortunato and Marc Barth\'{e}lemy.
\newblock Resolution limit in community detection.
\newblock {\em PNAS}, 104(1):36--41, January 2007.

\bibitem{mleval-pakdd}
Shantanu Godbole and Sunita Sarawagi.
\newblock Discriminative methods for multi-labeled classification.
\newblock In {\em PAKDD}, pages 22--30, 2004.

\bibitem{gurumine}
Amit Goyal, Byung-Won On, Francesco Bonchi, and Laks V.~S. Lakshmanan.
\newblock Gurumine: A pattern mining system for discovering leaders and tribes.
\newblock {\em ICDE}, 0:1471--1474, 2009.

\bibitem{hybrid}
Keith Henderson, Tina Eliassi-Rad, Spiros Papadimitriou, and Christos
  Faloutsos.
\newblock Hcdf: A hybrid community discovery framework.
\newblock In {\em SDM}, pages 754--765, 2010.

\bibitem{sampling}
Liran Katzir, Edo Liberty, and Oren Somekh.
\newblock Estimating sizes of social networks via biased sampling.
\newblock In {\em WWW}, pages 597--606, 2011.

\bibitem{lanci-comparative}
A.~Lancichinetti and S.~Fortunato.
\newblock Community detection algorithms: A comparative analysis.
\newblock {\em Physical Review E}, 80(5):056117--+, November 2009.

\bibitem{viralm}
Jure Leskovec, Lada~A. Adamic, and Bernardo~A. Huberman.
\newblock The dynamics of viral marketing.
\newblock {\em ACM Trans. Web}, 1, May 2007.

\bibitem{onnela}
Peter~J. Mucha, Thomas Richardson, Kevin Macon, Mason~A. Porter, and J-P
  Onnela.
\newblock Community structure in {Time-Dependent}, multiscale, and multiplex
  networks.
\newblock {\em Science}, 328(5980):876--878, 2010.

\bibitem{modularity}
M.~E.~J. Newman.
\newblock {Modularity and community structure in networks}.
\newblock {\em Proceedings of the National Academy of Sciences},
  103(23):8577--8582, June 2006.

\bibitem{cct}
Spiros Papadimitriou, Jimeng Sun, Christos Faloutsos, and Philip~S. Yu.
\newblock Hierarchical, parameter-free community discovery.
\newblock In {\em ECML PKDD}, pages 170--187, 2008.

\bibitem{walktrap}
Pascal Pons and Matthieu Latapy.
\newblock Computing communities in large networks using random walks.
\newblock {\em J. Graph Algorithms Appl.}, 10(2):191--218, 2006.

\bibitem{lp}
Usha~N. Raghavan, R\'{e}ka Albert, and Soundar Kumara.
\newblock {Near linear time algorithm to detect community structures in
  large-scale networks}.
\newblock {\em Physical Review E}, 76(3):036106+, September 2007.

\bibitem{infomap}
Martin Rosvall and Carl~T. Bergstrom.
\newblock {Maps of random walks on complex networks reveal community
  structure}.
\newblock {\em PNAS}, 105(4):1118--1123, January 2008.

\bibitem{bionet}
Jianhua Ruan and Weixiong Zhang.
\newblock An efficient spectral algorithm for network community discovery and
  its applications to biological and social networks.
\newblock {\em Data Mining, IEEE International Conference on}, 0:643--648,
  2007.

\bibitem{brl}
G.~Tsoumakas and I.~Katakis.
\newblock Multi label classification: An overview.
\newblock {\em International Journal of Data Warehousing and Mining},
  3(3):1--13, 2007.

\bibitem{dashun}
Dashun Wang, Zhen Wen, Hanghang Tong, Ching-Yung Lin, Chaoming Song, and
  Albert-L{\'a}szl{\'o} Barab{\'a}si.
\newblock Information spreading in context.
\newblock In {\em WWW}, pages 735--744, 2011.

\end{thebibliography}

\end{document}